\documentclass{aa}
\usepackage{graphicx}
\usepackage{txfonts}
\usepackage{natbib}

\newcommand{\Ha}{H$_{\alpha}$}
\newcommand{\Hb}{H$_{\beta}$}
\newcommand{\NII}{[NII]}
\newcommand{\EWmin}{EW$_{min}$}

\begin{document}
\title{Optical spectroscopy of BL Lacertae objects. Broad lines, companion 
galaxies and redshift lower limits.}
\author{B. Sbarufatti\inst{1}
       \and
       R. Falomo\inst{2}
       \and
       A. Treves\inst{1}
       \and
       J. Kotilainen\inst{3}}
\institute{Universit\`a dell'Insubria, Via Valleggio 11, I-22100 Como, Italy
       \and
           INAF, Osservatorio Astronomico di Padova, Vicolo dell'Osservatorio
           5, I-35122 Padova, Italy
	\and
	   Tuorla Observatory, University of Turku, V\"ais\"al\"antie 20, FIN-21500
            Piikki\"o, Finland}
\date{Received  / Accepted }

\abstract
{}
{We present optical spectroscopy of a sample of BL Lac objects, to determine 
their redshift, to study their broad emission line properties and to 
characterize their close environment.}
{Twelve  objects were observed using the  ESO 3.6m and 
the NOT 2.5m telescopes, obtaining spectra for the BL Lacs and for 
nearby sources.}
{For seven objects, nuclear emission lines and/or absorption lines from 
the host galaxy were detected. In all the four cases where absorption lines 
were revealed, the host galaxy has been resolved with HST or 
ground-based imaging. The broad  \Ha \ luminosities (or their upper limits) 
of the BL Lacs are similar to those of radio-loud quasars. 
For two BL Lacs, spectroscopy of close companions indicates that they are at 
the redshift of the BL Lacs, and therefore physically associated and likely 
interacting. 
Five BL Lacs have a featureless spectrum. In these cases, we apply a new 
technique to derive lower limits for their redshift. which are consistent with 
lower limits deduced from imaging. 
}
{}


\maketitle

\section{Introduction.}

BL Lacertae objects (hereafter BL Lacs) are a relatively rare 
subclass of active galactic nuclei (AGN) characterized by luminous, 
rapidly variable UV--to--NIR non--thermal continuum emission and polarization, 
strong compact flat spectrum radio emission and superluminal motion. 
Similar properties are observed also in flat spectrum radio quasars
and these two types of AGN are often grouped together into 
the class of blazars. The continuum emission of BL Lacs is boosted by 
relativistic beaming, that depresses the equivalent width (EW) 
of the spectral lines. 
However, apart from possible fluorescent emission lines, 
as in other types of AGN, absorption lines from 
the stellar population of the host galaxy, from intervening material and from 
the halo of the Milky Way are expected to be present in BL Lacs. 
These lines are probes of the physical conditions in the source, 
and in the intervening medium. They are obviously the most direct way of 
determining or constraining the redshift of the source.

The detection of weak lines requires high S/N spectroscopy that, 
for most BL Lacs, translates into a necessity to use large telescopes. 
Until recently, most of the work in this field have made use of 2-4 m 
class telescopes 
\citep[e.g.][]{falomo93, stickel93, veron93, bade94, falomo94, falomo96, 
marcha96, drinkwater97, rgb, landt01, rector01, londish02, hook03, C03}. 

However, significantly better results have been obtained with 8m 
class telescopes \citep[e.g.][]{heidt04,sowards05,woo05}. 
In particular, a substantial step forward in the detection of 
faint spectral lines was achieved by our extensive study of 42 BL Lacs 
performed with the ESO VLT \citep[][hereafter S05a and S06]{SI, SII}, 
where we determined the redshift for 18 sources, and developed a technique 
to obtain redshift lower limits for lineless sources.

In this paper, we complement the ESO-VLT dataset by observations with 2-4m 
class telescopes of sources that were not observed with the VLT, or that were 
observed in a different spectral range. The sample of 12 objects was taken 
from the list of BL Lacs in \citet{padovani95}, 
selecting the brightest targets 
among those with no available redshift, and bright, nearby targets to 
search for broad emission lines (in particular \Ha), and/or to study 
their environment. 
The first results of this campaign, concerning new redshifts, 
were published by \citet{C03}. Here we discuss the featureless objects, 
the search for broad lines and the properties of companions of the BL Lacs. 
For the five lineless objects we apply a technique, developed in S06, 
that allows us to set lower limits to the redshift.  
For the four low redshift objects we report a measurement or an upper limit 
for the broad component of the \Ha \ emission line.
In the case of \object{1H 0323+022} and \object{1ES 1440+122} we also present 
and discuss the spectra of their close companions. 
Throughout this paper, we adopt the following cosmological parameters: 
H$_0$= 70 km s$^{-1}$ Mpc$^{-1}$, $\Omega_{\Lambda}$=0.7, $\Omega_m$=0.3.

\section{Observations, data reduction and analysis.}

\begin{table*}[htbp]
  \caption{Journal of observations.}\label{tab:journal}
  \centering
  \begin{tabular}{lllclcllll}
  \hline
  \hline
\multicolumn{1}{c}{Object}& 
\multicolumn{1}{c}{RA}&
\multicolumn{1}{c}{Dec}&
\multicolumn{1}{c}{$z$}&
\multicolumn{1}{c}{Ref.$^1$}&
\multicolumn{1}{c}{{m$_R$}$^2$}&
\multicolumn{1}{c}{Setup}&
\multicolumn{1}{c}{S/N}&
\multicolumn{1}{c}{Exposure time}&
\multicolumn{1}{c}{Date}\\
\multicolumn{1}{c}{name}& 
\multicolumn{1}{c}{(J2000)}&
\multicolumn{1}{c}{(J2000)}&
\multicolumn{1}{c}{}&
\multicolumn{1}{c}{}&
\multicolumn{1}{c}{}&
\multicolumn{1}{c}{}&
\multicolumn{1}{c}{}&
\multicolumn{1}{c}{(s)}&
\multicolumn{1}{c}{}\\
\hline
\object{PKS 0109+224}	        &01 12 05  	&$+$22 44 54	&  *  		&	&15.6   &B      &230    &1800  	&26 Jul 01\\
\object{1H 0323+022}     	&03 26 13  	&$+$02 25 15	&0.147		& F86	&18.4   &A    	&30 	&2100	&15 Jan 02\\
     				&          	&           	&     		&    	&18.4   &C    	&30	&2400	&13 Jan 02\\
\object{PKS 0521-36}    	&05 22 57  	&$-$36 27 03	&0.055		& D79	&15.1   &C      &90     &1800  	&26 Jul 01\\
\object{PKS 0548-322}    	&05 50 41  	&$-$32 16 11	&0.068		& F76	&16.4   &C      &50     &1800  	&13 Jan 02\\
\object{MS0622.5-5256}   	&06 23 37  	&$-$52 57 57	&  *  		&	&19.2   &A      &20     &2400  	&14 Jan 02\\
\object{1ES 1106+244}		&11 09 16 	&$+$24 11 20 	&(0.46)$^3$	& S05b	&18.7   &N      &15     &3600  	& 9 Mar 05\\
\object{1ES 1239+069}		&12 41 48	&$+$06 36 01	&  *		&	&19.2   &A      &20     &2400  	&25 Jul 01\\
\object{PG 1437+398}		&14 39 18 	&$+$39 32 42 	&(0.26)$^3$	& S05b	&16.9   &N      &60     &3600	& 9 Mar 05\\
\object{1ES 1440+122}     	&14 42 48  	&$+$12 00 40	&0.162		& S93	&17.2   &A      &60     &2400	&24 Jul 01\\
\object{PKS 2005-489}    	&20 09 25  	&$-$48 49 54	&0.071		& F87	&13.9   &A      &400    &2400	&25 Jul 01\\
     				&          	&           	&     		&    	&13.9   &C	&250	&2400	&25 Jul 01\\
\object{PKS 2155-304}    	&21 58 51  	&$-$30 13 31	&0.116		& F93	&13.1   &A	&90     &2400	&25 Jul 01\\
\object{PKS 2201+04}		&22 04 18	&$+$04 40 02	&0.027		& V93	&15.2	&C	&100	&1800	&13 Jan 02\\
\hline
\multicolumn{10}{l}{\scriptsize $^1$: References for redshift determination:}\\
\multicolumn{1}{l}{}&\multicolumn{2}{l}{\scriptsize F76: \citet{fosbury76}}	&\multicolumn{3}{l}{\scriptsize F87: \citet{falomo87}}	&\multicolumn{4}{l}{\scriptsize V93: \citet{veron93}}\\
\multicolumn{1}{l}{}&\multicolumn{2}{l}{\scriptsize D79: \citet{danziger79}}	&\multicolumn{3}{l}{\scriptsize S93: \citet{schachter93}}&\multicolumn{4}{l}{\scriptsize S05b: \citet{SIII}}\\
\multicolumn{1}{l}{}&\multicolumn{2}{l}{\scriptsize F86: \citet{filippenko86}}	&\multicolumn{3}{l}{\scriptsize F93: \citet{falomo93}}  &\multicolumn{4}{l}{\scriptsize }\\
\multicolumn{10}{l}{\scriptsize $^2$: R-band magnitudes determined from the flux measured in our spectra.}\\
\multicolumn{10}{l}{\scriptsize $^3$: The imaging redshift estimate from S05b.}\\
\end{tabular}
\end{table*}

In Table \ref{tab:journal} we report a journal of the observations and in 
Table \ref{tab:setup} the instrumental configurations. 
The low resolution setups (A, B, N) were used for the redshift determination 
and for studying sources in the close environment of the BL Lacs, 
while the medium resolution setup (C) was used for the search for 
the broad components of emission lines, especially \Ha.

\begin{table}[htb]
\caption{Instrumental configurations.}\label{tab:setup}
\scriptsize
\begin{tabular}{cccccr}
\hline
\hline
\multicolumn{1}{c}{Setup}&
\multicolumn{1}{c}{Telescope}&
\multicolumn{1}{c}{Slit}&
\multicolumn{1}{c}{Range}&
\multicolumn{1}{c}{Disp.}&
\multicolumn{1}{c}{Resol.}\vspace{-3pt}\\
\multicolumn{1}{c}{}&
\multicolumn{1}{c}{}&
\multicolumn{1}{c}{\scriptsize (arcsec)}&
\multicolumn{1}{c}{\scriptsize (\AA)}&
\multicolumn{1}{c}{\scriptsize (\AA \, pix$^{-1}$)}&
\multicolumn{1}{c}{\scriptsize (\AA)}\\
\hline
\textbf{A}&ESO 3.6 + EFOSC2& 1.2&4085--7520&1.68&10.5\\
\textbf{B}&ESO 3.6 + EFOSC2& 1.2&5200--9350&2.06&12.8\\
\textbf{C}&ESO 3.6 + EFOSC2& 1.0&6285--8210&0.95&5.9 \\
\textbf{N}&NOT 2.5 + ALFOSC& 2.5&3200--9100&3.00&32.4 \\
\hline
\end{tabular}
\end{table}

Data reduction was performed using IRAF\footnote{IRAF (Image Reduction and 
Analysis Facility) is distributed by the National Optical Astronomy 
Observatories, which are operated by the Association of Universities for 
Research in Astronomy, Inc., under cooperative agreement with the National 
Science Foundation.} \citep{tody86,tody93}, following standard procedures 
for spectral analysis. This includes bias subtraction, flat fielding and  
removal of bad pixels. For each target, we obtained two spectra in order to 
get a good correction of the cosmic rays and to check for the reality of 
weak spectral features. The individual frames were then combined into 
a single average image. 
Wavelength calibration was performed using the spectra of 
a Helium/Neon/Argon lamp obtained during the same observing night, 
resulting in an accuracy of $\sim$3 \AA \ (rms). 
From these calibrated final images we extracted the one-dimensional spectra 
adopting an optimal extraction algorithm \citep{horne86} to improve the S/N. 

Although this program did not require optimal photometric conditions, 
the sky was clear during most of the observations. This enables us to perform 
a spectrophotometric calibration of the data using standard stars 
\citep{oke90} observed during the same nights. 
The ESO spectral setups B and C at wavelengths $\lambda>$7000 \AA \  
are affected by fringing. This was corrected for using flat field images 
taken immediately before or after the spectrum of the object, 
with the telescope in the same position. 
In the cases where such flat field images were not available, 
the resulting fringing pattern makes the detection of features 
in this spectral range very difficult, if not impossible. 
Finally, all the spectra were dereddened following the extinction law by 
\citet{cardelli89}, and assuming the E$_{\textrm{B-V}}$ values computed by 
\citet{schlegel98}. 

In Fig. \ref{fig:esospectra}, we present the optical spectrum of each source. 
In order to show more clearly the continuum shape and the faint features, 
we report both the flux calibrated and the normalized spectrum 
for each object. The main emission and absorption features are identified, 
interstellar absorption features are marked as ISM, and telluric absorption 
features as $\oplus$.
All these spectra are electronically available in our spectroscopic library of 
BL Lacs, at \emph{http://www.oapd.inaf.it/zbllac/}.

\subsection{Redshift lower limits.}

For five objects the spectra appear featureless. In these cases, 
using the minimum observable EW (\EWmin), it is possible to derive 
a lower limit for the redshift.
This procedure is described in detail in S06. Briefly, it is based on 
the assumption that the host galaxy is an elliptical 
with an absolute magnitude of M$_R\simeq-22.9\pm0.5$, as derived from 
the analysis of a homogeneous sample of HST images of BL Lacs 
\citep[see ][hereafter S05b]{urry00,SIII}. 
Adopting a template for the spectrum of an elliptical galaxy \citep{kinney96}, 
it was shown by S06 that from the apparent magnitude of the BL Lac, 
and the \EWmin in the spectrum, one can obtain a lower limit to the redshift. 
The lower limits from this procedure and their comparison with redshifts 
or lower limits deduced from the imaging of the host galaxy are reported in 
Table \ref{tab:featureless}.

\begin{table}[htbp]
\centering
\caption{Featureless objects.\label{tab:featureless}}
\begin{tabular}{lllr}
\hline
\hline
\multicolumn{1}{c}{Object name}&\multicolumn{1}{c}{\EWmin}&\multicolumn{1}{c}{$z_{min}$}&\multicolumn{1}{r}{${z_{ima}}^1$}\\
\multicolumn{1}{c}{}&\multicolumn{1}{c}{}&\multicolumn{1}{c}{}&\multicolumn{1}{c}{}\\
\hline
\object{PKS 0109+224}	& 0.43	& $>$0.18 & $>$0.40	\\
\object{MS 0622.5-525}	& 1.48	& $>$0.49 & $>$0.41	\\
\object{1ES 1106+244}	& 2.5	& $>$0.29 & 0.46	\\
\object{1ES 1239+069}	& 0.75	& $>$0.60 & $>$0.92	\\
\object{PG 1437+398}	& 0.8   & $>$0.24 & 0.26	\\
\hline
\multicolumn{4}{l}{\scriptsize $^1$: Imaging redshifts from S05b. 
For PKS  0109+224, see \citet{falomo96}.}\\
\end{tabular}
\end{table}

\section{Results for individual objects.}

\paragraph{\object{PKS 0109+224}}: This radio source was discovered in 
the 5 GHz Green Bank survey \citep{davis71} and subsequently classified as 
a BL Lac by \citet{owen77}. 
It exhibits significant variability in flux and polarization in both radio and 
optical bands \citep{ciprini04}.
The host galaxy was not detected in images obtained at 
the NTT \citep{falomo96} and the NOT \citep{nilsson03}. 
The claimed detection of the host by \citet{wright98} (m$_K$=12.2) is dubious, 
since the host is resolved only in one of their two images. 
Based on the non-detection of the host and assuming that the host has 
an absolute magnitude M$_R$=-23.5, \citet{falomo96} proposed a lower limit to 
its redshift of $z>$0.4. 
Previous low-medium S/N optical spectra were featureless 
\citep{wills79,falomo94}. Although we reach a high S/N (S/N = 230) in the red 
(5500--9000 \AA), the spectrum remains featureless (Fig. \ref{fig:esospectra}).
We determine \EWmin=0.43 \AA \ which according to our adopted procedure 
implies a redshift lower limit of $z>$0.18, consistent with, but considerably 
lower than the limit from imaging.

\paragraph{\object{1H 0323+022}}: This X-ray selected object \citep{doxsey83} 
was classified as a BL Lac by \citet{margon84}. It exhibits significant 
optical polarization \citep{feigelson86} and variability \citep{villata00}. 
The host galaxy was resolved in ground-based images by \citet{feigelson86}, 
\citet{falomo96} and \citet{nilsson03}. 
The signature of the host galaxy at $z$=0.147 was also clearly apparent in 
the optical spectrum of \citet{filippenko86}. 
Our new spectrum (Fig. \ref{fig:esospectra}), despite being dominated by 
the strong nuclear component, clearly shows several absorption lines from 
the host galaxy, i.e. CaII $\lambda\lambda$3934,3968 and G band $\lambda$4305 
(setup A), and NaI $\lambda$5892 (setup C), 
confirming its redshift as $z$=0.147. 
The spectrum also reveals the narrow [NII] $\lambda$6583 emission line at 
this redshift. However, no broad \Ha \ emission is detected, 
with an \EWmin \ limit of 1 \AA, corresponding to an upper limit of 
1.8 10$^{40}$ erg s$^{-1}$ for the broad \Ha \ luminosity.

This BL Lac is located in a complex environment \citep[e.g.][]{falomo96}, 
as shown in Fig. \ref{fig:0323ima}. At a distance of $\sim$1' east of 
the BL Lac there is a bright elliptical galaxy (G1) at a similar redshift as 
the BL Lac \citep[$z\sim$0.16,][]{falomo96}. 
In the close environment of the BL Lac, there are a number of complex 
emission features (Fig. \ref{fig:0323ima}). In particular, 
a compact knot-like structure (G2) is located at a distance of $\sim$2.6''. 
We observed \object{1H 0323+022} using the setup A with the slit slightly 
shifted with respect to the nucleus, to simultaneously obtain also the spectra 
of G1 and G2 (Fig. \ref{fig:0323spc}).  
G1 is a bright elliptical galaxy at a redshift of $z$=0.160, determined from 
the CaII H \& K, G band, and MgI absorption lines. 
The spectrum of G2 (Fig. \ref{fig:0323spc}, second panel), because of its 
small angular distance from the BL Lac and its low surface brightness, 
is contaminated by the light from the BL Lac and its host galaxy. Therefore we 
extracted a spectrum using 
an identical aperture size to the one used for G2, taken in a position 
symmetric to G2 with respect to the position of the BL Lac 
(Fig. \ref{fig:0323spc}, third panel), and subtracted it 
from the spectrum of G2. 
The decontaminated spectrum of G2 (Fig. \ref{fig:0323spc}, bottom panel) 
has the characteristic shape of an elliptical galaxy, 
with the absorption features of CaII $\lambda\lambda$3934,3968, G band
$\lambda$4305 and MgI $\lambda$5175 at $z$=0.148. 
The measured flux from the spectrum leads to an estimate of R$\simeq$18.6, 
which corresponds to M$_R$=-20.9. Therefore, G2 could be an elliptical 
dwarf galaxy at the redshift of \object{1H 0323+022}, 
as already suggested by  \citet{falomo96}. The projected distance of G2 from 
the BL Lac is only $d\simeq$8 kpc.

\paragraph{\object{PKS 0521-36}}: This is a well studied BL Lac which has been 
observed extensively at all wavelengths 
\citep[see e.g.][and references therein]{pian96}. 
The host galaxy has been resolved in several imaging studies 
\citep[e.g.][]{wurtz96,falomo94,kotilainen98,urry00,cheung03}. 
The redshift of this BL Lac ($z$=0.055) \citet{danziger79} is based on 
both absorption lines from the host galaxy and strong emission lines from 
the nucleus. In particular, a broad (FWHM$\sim$3000 km s$^{-1}$) \Ha \ 
with varying EW (ranging from 20 to 80 \AA) has been detected 
\citep{falomo94}. 
Our high S/N spectrum (Fig. \ref{fig:esospectra}) 
clearly shows the narrow [OII] $\lambda$6300, 
HeI $\lambda$5875 and SII $\lambda\lambda$6716,6730 emission lines, 
and the broad \Ha+[NII] blend, with EW=40.7 \AA, corresponding to 
a broad \Ha \ luminosity of 7.5 10$^{41}$ erg s$^{-1}$, 
which is within the range previously observed by \citet{falomo94}.

\paragraph{\object{PKS 0548-322}}: This X-ray selected BL Lac at $z$=0.068 
\citep{fosbury76} is located in a rich environment \citep{falomo95}. 
At least one of the companions shows signs of interaction with the BL Lac. 
The host galaxy has been detected both in imaging \citep{urry00} 
and in spectroscopy by \citet{falomo00}, who did not reveal emission lines.
Our new spectrum (Fig. \ref{fig:esospectra}) clearly shows 
a narrow emission line which we identify as [NII]$\lambda$6583 at z=0.068. 
The presence of [NII] emission could be a signature of recent star formation 
in the host galaxy, induced by the interaction with the close companion 
\citet{falomo95}. No other emission features are detected, 
and the \EWmin \ limit of 0.9 \AA \ corresponds to an upper limit of 
8.4 10$^{39}$ erg s$^{-1}$ for the broad \Ha \ luminosity, i.e. five times 
smaller than the upper limit by \citet{falomo00}.

\paragraph{\object{MS0622.5-5256}}: This is an X-ray selected BL Lac belonging 
to the EINSTEIN Extended Medium Sensitivity Survey \citep[EMSS,][]{gioia90}.  
Previous low S/N spectroscopy \citep{stocke85} showed a featureless continuum. 
HST imaging of this BL Lac \citep{scarpa00a} failed to resolve 
the host galaxy, suggesting a redshift lower limit $z>$0.4 (see S05b). 
Our new spectrum (Fig. \ref{fig:esospectra}) is featureless, 
with \EWmin=1.48 \AA , which gives a spectroscopic redshift lower limit of 
$z>$0.49, well consistent with the lower limit from imaging.

\paragraph{\object{1ES 1106+244}}: This BL Lac belongs to 
the EINSTEIN Extended Medium Sensitivity Survey \citep[EMSS,][]{gioia90}. 
The host galaxy has been resolved both in ground-based \citep{falomo99} 
and in HST imaging \citep{urry00}, indicating a redshift lower limit 
of $z\simeq$0.42. Earlier spectroscopy of this BL Lac by \citet{perlman96} 
showed a featureless spectrum. From our moderate S/N spectrum 
(Fig. \ref{fig:esospectra}) we determine \EWmin=2.50 \AA, which implies 
a redshift lower limit of $z>0.29$, consistent with the limit obtained 
from imaging.

\paragraph{\object{1ES 1239+069}}: This high energy peaked BL Lac (HBL) 
has been proposed as a candidate TeV source by \citet{stecker96}, 
being supposedly at a relatively low redshift 
\citep[$z$=0.150, ][]{perlman96}. However, this redshift estimate, 
based on the possible detection of absorption features from the host galaxy, 
is ruled out by our new spectrum (Fig. \ref{fig:esospectra}), from which 
the measured \EWmin = 0.75 \AA ~implies $z>$0.60. 
Moreover, the non-detection of the host galaxy in imaging sets 
a further lower limit of $z>$0.92 (S05b), considerably higher than 
the spectroscopic one, making the detection of this BL Lac 
in the TeV domain unlikely. Indeed \citet{horan04} failed to detect 
this BL Lac using the Whipple 10 m $\gamma$-ray telescope.

\paragraph{\object{PG 1437+398}}: This HBL belongs to the Sedentary Survey 
\citep{SS}. Its host galaxy was resolved with HST imaging, 
giving an imaging redshift of $z\simeq$0.26 (S05b). 
Previous optical spectroscopy \citep{rgb,white00,scarpa95} have led to 
a featureless spectrum. Note that the redshift $z$=0.34 reported for this 
BL Lac by the NASA Extragalactic Database is  based on a very low S/N 
($\sim$5) Sloan Digital Sky Survey 
spectrum\footnote{see Sloan Digital Sky Survey Data Release 4, 
(http://cas.sdss.org/astro/en/tools/getimg/spectra.asp), plate 
1350/52786, fiber 333, and \citet{richards02} for a description of the
quasar survey}. This redshift is ruled out by our new, much higher S/N 
(S/N $\sim$60) spectrum (Fig. \ref{fig:esospectra}), 
which shows a featureless continuum. From the \EWmin \ value (\EWmin = 0.8 \AA)
we deduce a redshift lower limit of $z>$0.24, 
consistent with the imaging redshift estimate.

\paragraph{\object{1ES 1440+122}}: This is an X-ray selected BL Lac belonging 
to the Einstein Slew Survey. It is located in a rich environment, 
(Fig. \ref{fig:1440ima}), 
being surrounded by $\sim$20 galaxies \citep{heidt99}, thus suggesting 
that this BL Lac is located in a group or a small cluster of galaxies. 
The host galaxy has been resolved in several imaging studies 
\citep{heidt99,falomo99,urry00,kotilainen04}. 
High resolution HST imaging by \citet{scarpa99} revealed 
a very close companion ($\sim$ 0.3'') to this BL Lac, 
suggesting the possibility of gravitational lensing. 
This hypothesis was, however, ruled out by a radio-optical study by 
\citet{giovannini04} who demonstrated that the companion object is 
a foreground star. 
We obtained spectra of the BL Lac itself (Fig. \ref{fig:esospectra}), 
and of a galaxy at a distance of 25'' (G1) and of a close companion 
at a distance of $\sim$2'' (G2; Fig. \ref{fig:1440spc}).
The spectrum of \object{1ES1440+122} is dominated by emission from 
the host galaxy, but the contribution of the nucleus becomes apparent 
towards blue wavelengths 
(indeed, the strength of the CaII break is only $\sim$20\%). 
The redshift of this BL Lac, $z$=0.162, 
measured from the CaII $\lambda\lambda$3934,3968, G band $\lambda$4305 and 
MgI $\lambda$5175 absorption lines from the host galaxy, 
confirms the result by \citet{schachter93}.
G1 is a typical elliptical galaxy at a redshift of $z$=0.164, 
and at a projected distance from the BL Lac of $\sim$98 kpc. The other 
companion, G2, is also an elliptical galaxy at a redshift of $z$=0.161 
\citep[see also][who discuss their unpublished spectrum of G2]{nilsson03}, 
and at a projected distance from the BL Lac of only $\sim$4 kpc. 
Such a small separation strongly indicates that the BL Lac and the galaxy G2 
are interacting.

\paragraph{\object{PKS 2005-489}}: This X-ray selected BL Lac is at 
low redshift \citep[z=0.071 by ][based on the detection of \Ha \ and \NII \ 
emission lines]{falomo87}. More recent spectroscopy has revealed 
weak NI $\lambda$1135 and CIII $\lambda$1176 emission lines from the nucleus 
\citep{penton04}, and absorption lines from the host galaxy \citep{pesce94}. 
The host galaxy of this BL Lac, resolved in several optical and NIR 
imaging studies \citep{stickel93, falomo96, kotilainen98, urry00, cheung03}, 
is a giant elliptical with M$_R$=-23.1, 
in a relatively rich environment \citep{pesce94}. Several of the nearby 
galaxies are known to be at the redshift of the BL Lac 
\citep{stickel93,pesce94}. The optical spectrum of this BL Lac is 
strongly dominated by the nuclear continuum 
\citep{falomo94, perlman96}. The very high S/N reached by our new spectrum 
($\sim$400, setup A; Fig. \ref{fig:esospectra}) allowed the detection of 
the spectroscopic signatures from the host galaxy. 
The CaII $\lambda\lambda$3934,3968, G band $\lambda$4305 and 
MgI $\lambda$5175 absorption lines have EW ranging from 0.2 to 0.4 \AA.
The spectrum obtained with setup C confirms the presence of \Ha \ 
$\lambda$6563 and [NII] $\lambda$6583 narrow emission lines. 
No broad component of \Ha is detected, with an \EWmin \ limit of 0.2 \AA, 
corresponding to an upper limit of 1.9 10 $^{41}$ erg s$^{-1}$ 
for the broad \Ha \ luminosity.

\paragraph{\object{PKS 2155-304}}: Although this HBL, a prototype of 
its class, has been studied in a large number of papers at all wavelengths 
\citep[e.g.][and references therein]{pesce97}, little optical spectroscopy 
has been published. Its redshift, $z$=0.116, has been measured by 
\citet{falomo93}, from the G band, MgI and NaI absorption lines 
(and a marginally detected CaII doublet) in a spectrum of the host galaxy with 
the slit offset from the nucleus. 
The host galaxy has been resolved in imaging studies, 
with I=14.8 \citep{falomo91,kotilainen98}, consistent with this redshift.
Our new high S/N ESO spectrum (Fig. \ref{fig:esospectra}), 
also taken with the slit offset from the nucleus, allows us to confirm 
the detection of all the features reported by \citet{falomo93}. In particular, 
the CaII lines $\lambda\lambda$3934,3968 are clearly revealed, 
with EW of 0.5 and 0.4 \AA \ for the K and H lines, respectively.

\paragraph{\object{PKS 2201+04}}: This radio source was classified as a BL Lac 
by \citet{weiler80}. Its redshift, $z$=0.026, proposed by \citet{wills76} 
has been confirmed by more recent observations, showing absorption lines from 
the host galaxy, along with narrow and broad nuclear emission lines, 
in particular, broad components of \Hb \ and \Ha
\citep{falomo87, veron93}.
Our new spectrum (Fig. \ref{fig:esospectra}) reveals the [OII] $\lambda$6300, 
[NII] $\lambda$6583, \Ha \ $\lambda$6563 \ and SII $\lambda$6730 
emission lines. The \Ha-[NII] blend clearly shows the presence of 
a broad component, with an EW of 13.8 \AA, corresponding to 
a broad \Ha \ luminosity of 6.1 10$^{40}$ erg s$^{-1}$.

\section{Conclusions.}

We have presented new, high quality optical spectroscopy of a sample of 
12 BL Lac objects. Absorption lines from the host galaxy were detected
in four objects. In all these cases, the host galaxy has also been resolved 
in imaging, either with HST \citep[\object{1ES 1440+122}, ][]{urry00} 
or from the ground \citep[\object{1H 0323+022}, \object{PKS 2005-489}, 
\object{PKS 2155-304}, see][ respectively]{feigelson86,pesce94,kotilainen98}. 
The absolute magnitude of the host is in all cases close to M$_R$= -22.9, 
which is typical for BL Lac hosts (S05b). The characterization of 
the absorption lines from the host galaxy is, however, arduous in most cases, 
because the EW of the lines is strongly reduced by 
the beamed non-thermal continuum.

The broad emission line intensities of BL Lacs are similar to those of 
radio-loud quasars \citep[e.g.][]{pian05}. The two cases where broad \Ha \ is 
observed in this study (\object{PKS 0521-365}, 
L$_{H_{\alpha}}\simeq$7.5 10$^{41}$ erg s$^{-1}$; 
\object{PKS 2201+04} L$_{H_{\alpha}}\simeq$6.1 10$^{40}$ erg s$^{-1}$) 
confirm this conclusion. It is also consistent with the derived upper limits 
in the cases of \object{1H 0323+022} 
( L$_{H_{\alpha}}<$1.8 10$^{40}$ erg s$^{-1}$), \object{PKS 0548-322} 
(L$_{H_{\alpha}}<$8 10$^{39}$ erg s$^{-1}$) and \object{PKS 2005-489} 
(L$_{H_{\alpha}}<$1.9 10$^{41}$ erg s$^{-1}$).

For two BL Lacs (\object{1H 0323+022} and \object{1ES 1440+122}) 
we have demonstrated that very nearby (projected distance 4--8 kpc) 
companion galaxies are at the redshift of the BL Lac, indicating that there 
is a physical association and a likely interaction. Similar cases have been
previously found for a number of other BL Lacs 
\citep[see][]{pesce95,falomo96,falomo00a}. 

Finally, we consider the five BL Lacs which remain featureless in our spectra. 
Their redshift lower limits, based on the minimum observable EW of the 
non-detected absorption lines, appear consistent with the ones deduced 
from imaging. As already noted by S05b, the imaging technique is 
more stringent for brighter objects, but the spectral technique is the only 
available method for faint (m$_V \geq$18) sources.

\begin{figure*}
  \centering
  \includegraphics[scale=0.8]{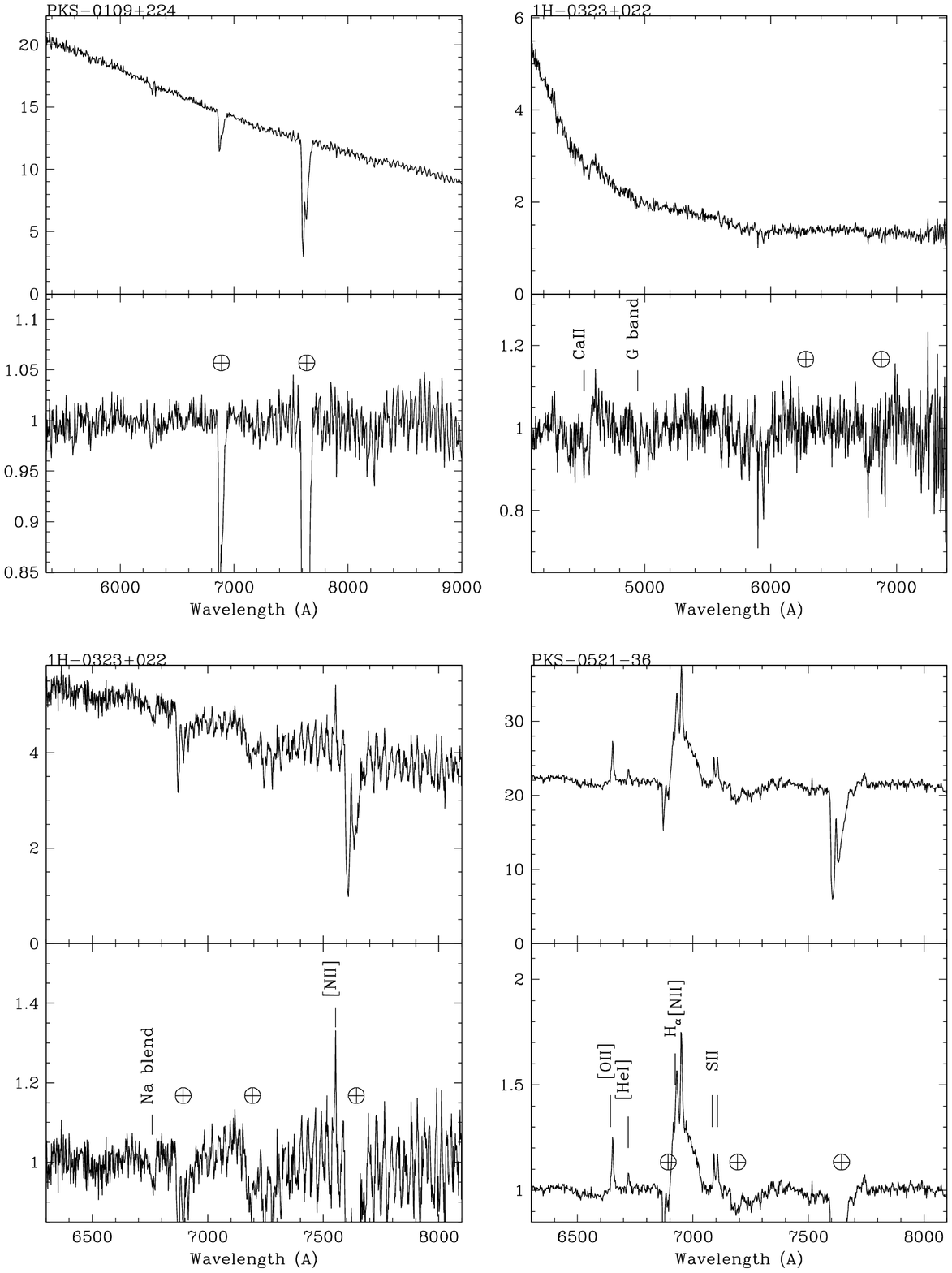}
  \caption{\scriptsize Spectra of the BL Lacs observed with the ESO 3.6m and 
the NOT 2.5m telescopes. Top panel: flux calibrated spectra. Bottom panel: 
spectra normalized with respect to the continuum.
   Telluric bands are indicated by $\oplus$, spectral lines are marked by the
   line ID, and absorption features from atomic species in 
the interstellar medium of our galaxy are labeled by ISM.}
\label{fig:esospectra}
\end{figure*}\addtocounter{figure}{-1}
\begin{figure*}
  \centering
  \includegraphics[scale=0.8]{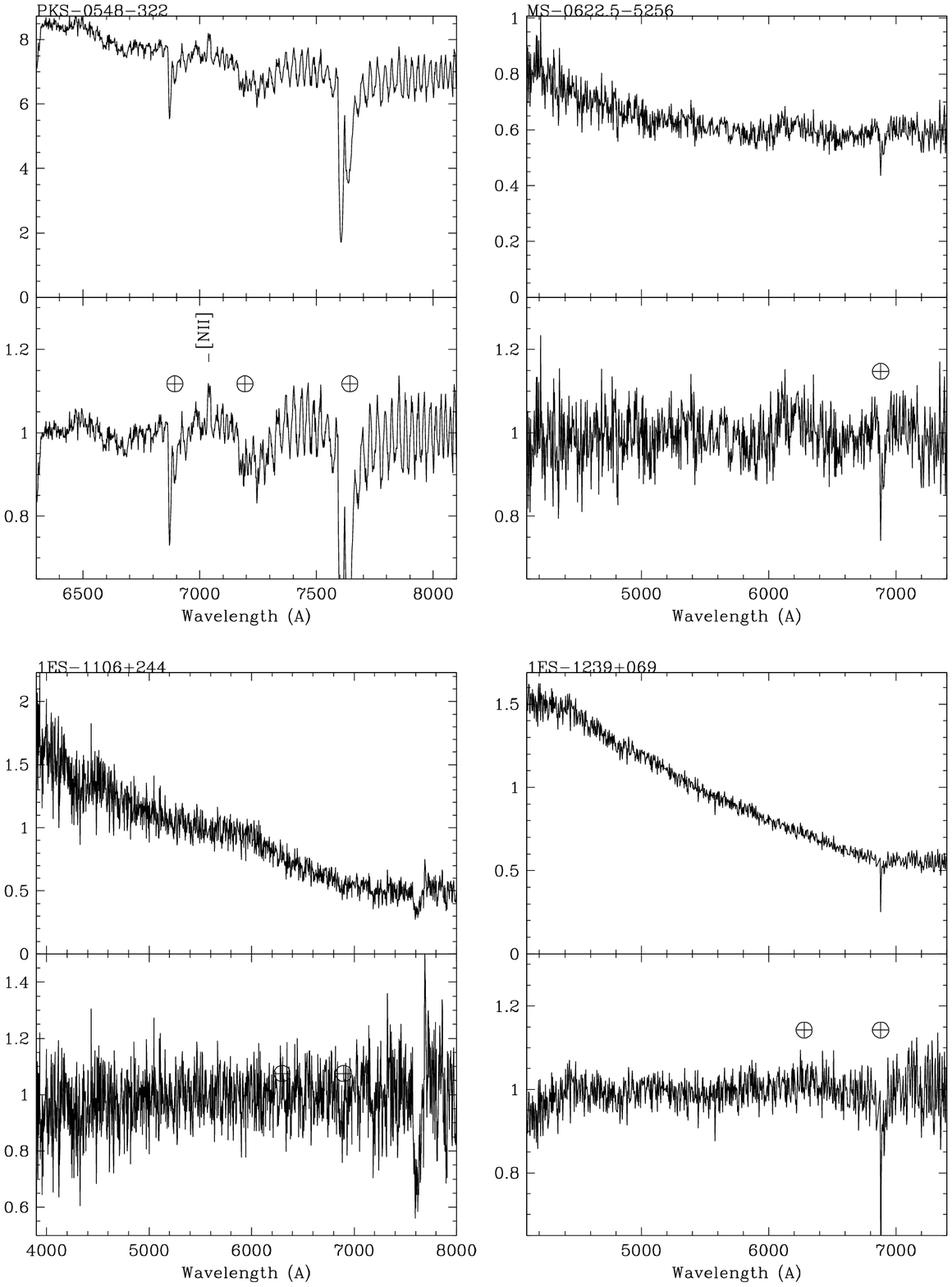}
  \caption{continued}
\end{figure*}\addtocounter{figure}{-1}
\begin{figure*}
  \centering
  \includegraphics[scale=0.8]{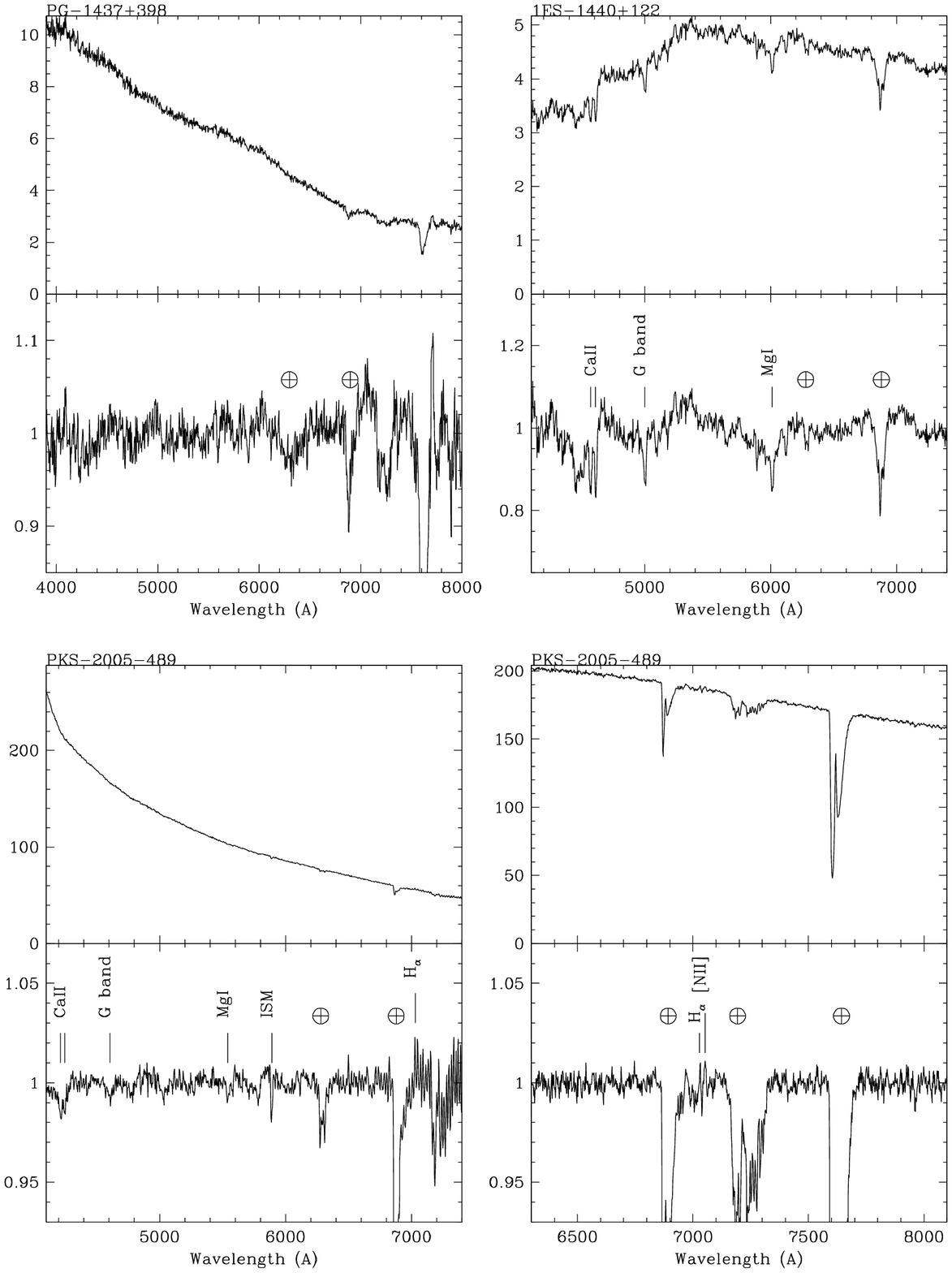}
  \caption{continued}
\end{figure*}\addtocounter{figure}{-1}
\begin{figure*}
  \centering
  \includegraphics[scale=0.8]{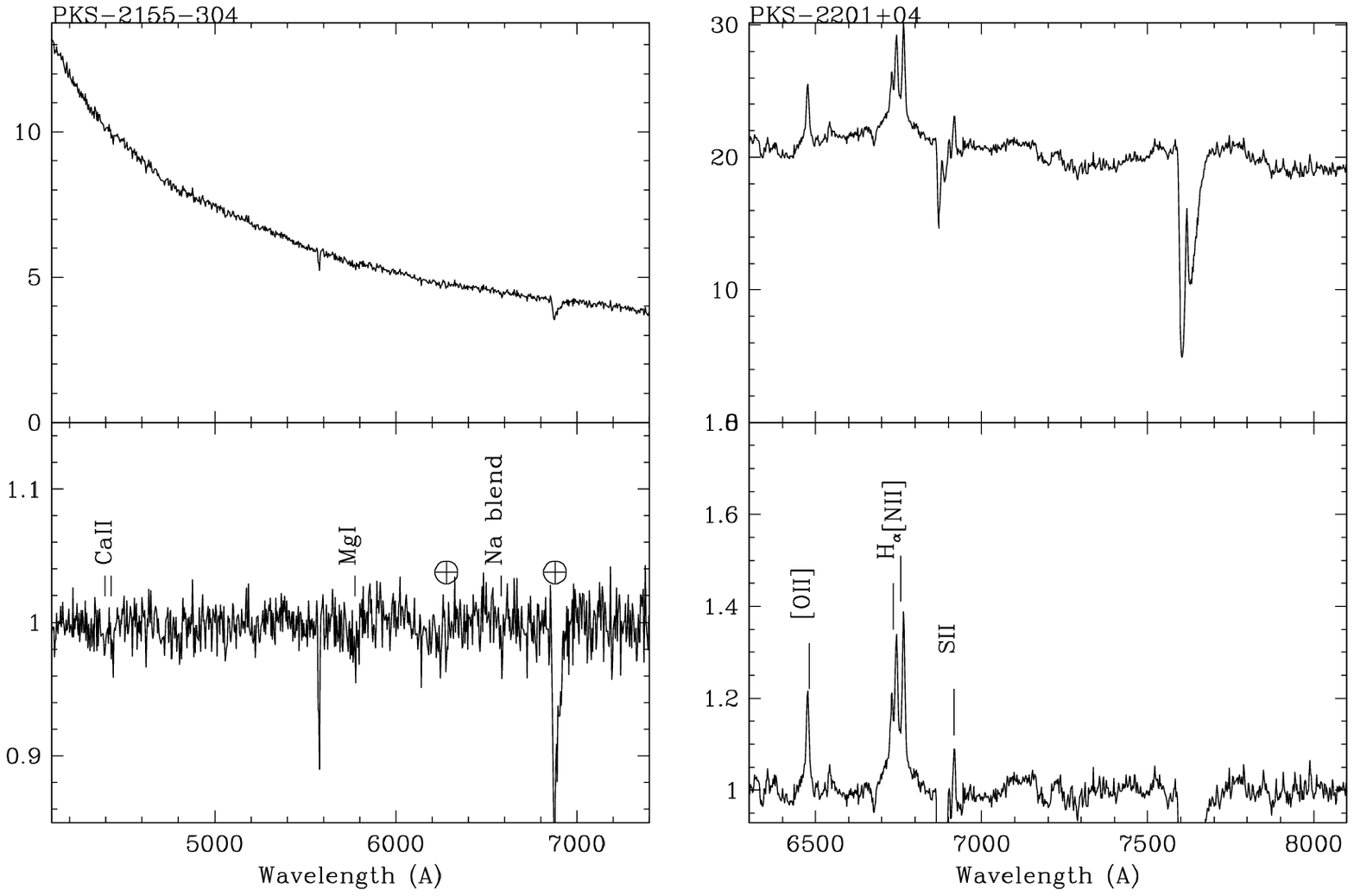}
  \caption{continued}
\end{figure*}

\begin{figure*}
\centering\includegraphics[scale=0.8]{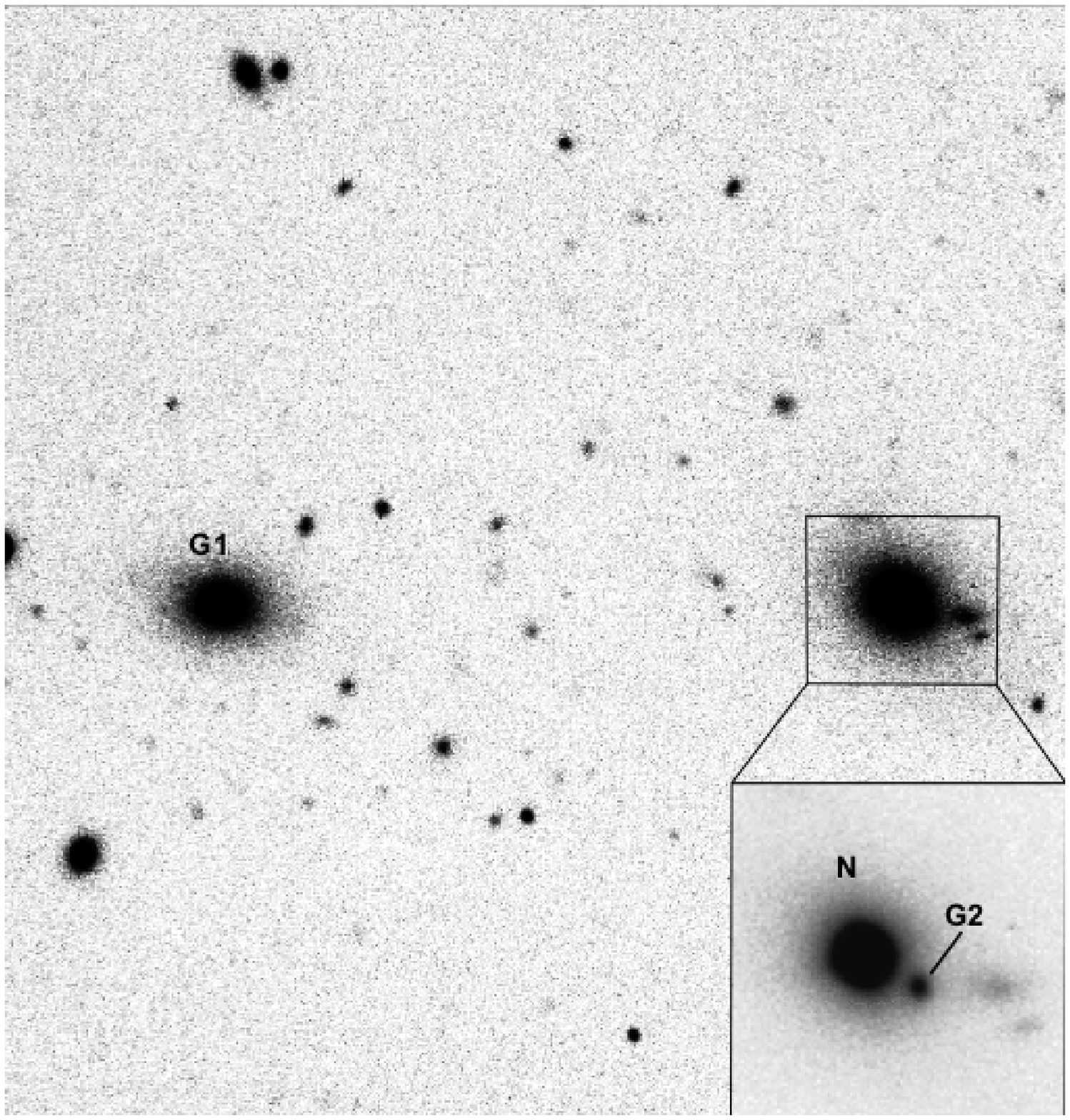}
\caption{The environment of \object{1H 0323+22}. In the main panel 
  (ESO 3.6m raw image, 10 s exposure), the
  companion galaxy G1 is labeled. The subpanel in the lower right hand corner 
(R-band image, adapted from  \citet{falomo96}) shows a
  higher resolution image of the BL Lac. The nucleus surrounded
  by the host galaxy is apparent together with the compact knot G2. The
  angular separation between the nucleus and G2 is 2.6''}\label{fig:0323ima}
\end{figure*} 

\begin{figure*}
\centering\includegraphics[scale=0.8]{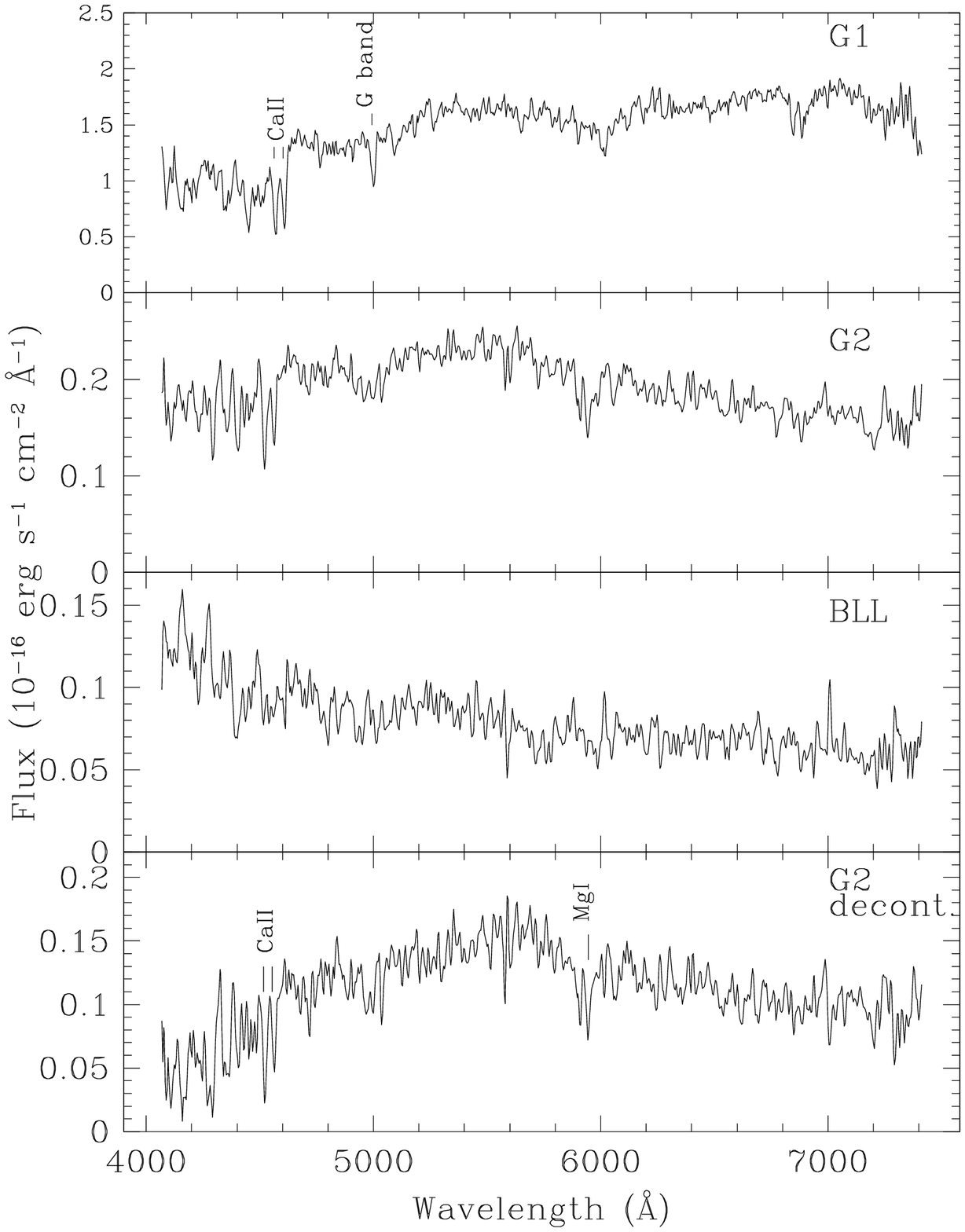}
\caption{Spectra of the companion objects of \object{1H 0323+22}. 
Top panel: G1;  Second panel: G2  (before decontamination);
  Third panel: Spectrum of the BL Lac, taken with an aperture symmetric to G2;
  Bottom panel: G2 (decontaminated spectrum)}\label{fig:0323spc}
\end{figure*}

\begin{figure*}
\centering\includegraphics[scale=0.9]{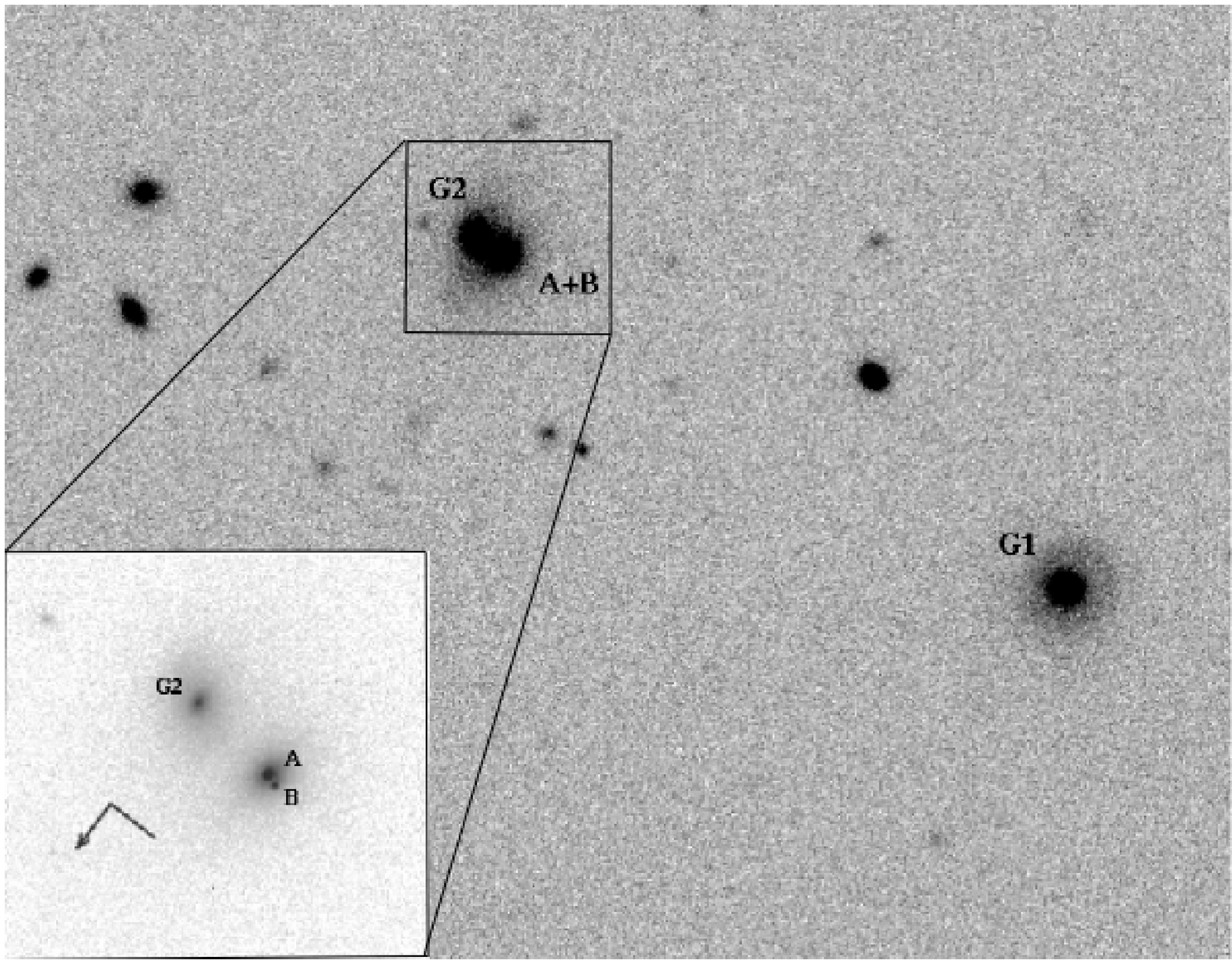}
\caption{The environment of \object{1ES 1440+122}. In the main panel 
(R-band image taken at the  NOT by R. Falomo), the locations of 
galaxies G1 and G2 and the nucleus (A+B)   are indicated. 
  Subpanel: HST detail of the region around the BL Lac
  \citep[filter F702W image from][]{scarpa99}; 
  A is the BL Lac nucleus, B is a foreground star  (see \citet{giovannini04}). 
The angular separation between A and G2 is 2''.}\label{fig:1440ima}
\end{figure*}

\begin{figure*}
\centering\includegraphics[scale=0.8]{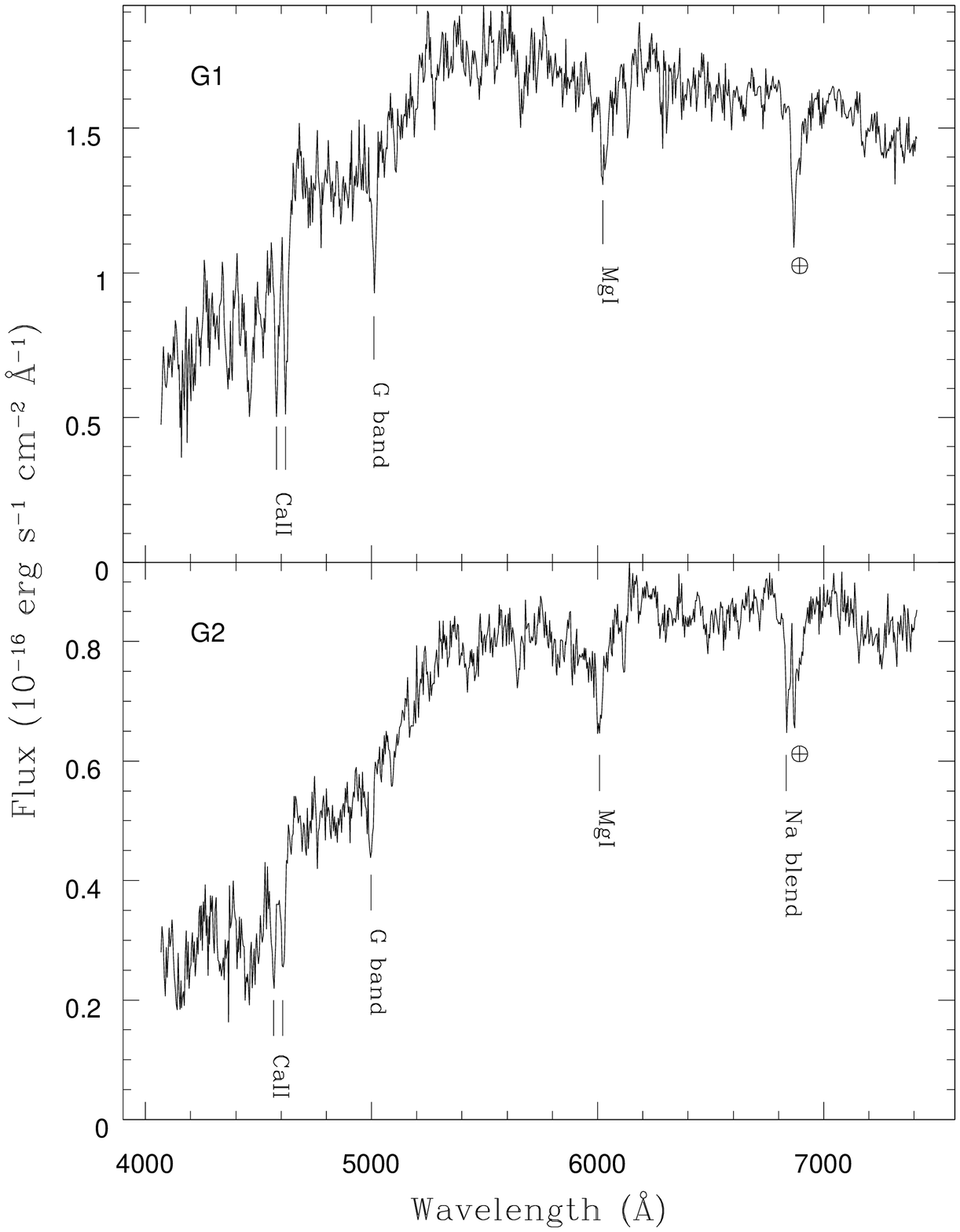}
\caption{Spectra of the companion galaxies G1 (top) and G2 (bottom) of 
\object{1ES 1440+122} 
}\label{fig:1440spc}
\end{figure*}


\begin{thebibliography}{}

\bibitem[Bade, Fink, \& Engels(1994)]{bade94} Bade, N., Fink, 
H.~H., \& Engels, D.\ 1994, \aap, 286, 381 

\bibitem[Carangelo et al.(2003)]{C03} Carangelo, N., 
Falomo, R., Kotilainen, J., Treves, A., \& Ulrich, M.-H.\ 2003, \aap, 412, 
651

\bibitem[Cardelli, Clayton, \& Mathis(1989)]{cardelli89} Cardelli, 
J.~A., Clayton, G.~C., \& Mathis, J.~S.\ 1989, \apj, 345, 245 

\bibitem[Cheung et al.(2003)]{cheung03} Cheung, C.~C., Urry, 
C.~M., Scarpa, R., \& Giavalisco, M.\ 2003, \apj, 599, 155 

\bibitem[Ciprini et al.(2004)]{ciprini04} Ciprini, S., Tosti, G., 
Ter{\"a}sranta, H., \& Aller, H.~D.\ 2004, \mnras, 348, 1379 

\bibitem[Danziger et al.(1979)]{danziger79} Danziger, I.~J., 
Fosbury, R.~A.~E., Goss, W.~M., \& Ekers, R.~D.\ 1979, \mnras, 188, 415 

\bibitem[Davis(1971)]{davis71} Davis, M.~M.\ 1971, \aj, 76, 980 

\bibitem[Doxsey et al.(1983)]{doxsey83} Doxsey, R., Bradt, H., 
McClintock, J., Petro, L., Remillard, R., Schwartz, D., Wood, K., \& 
Ricker, G.\ 1983, \apjl, 264, L43 
 
\bibitem[Drinkwater et al.(1997)]{drinkwater97} Drinkwater, M.~J., 
et al.\ 1997, \mnras, 284, 85 

\bibitem[Falomo et al.(1987)]{falomo87} Falomo, R., Maraschi, 
L., Treves, A., \& Tanzi, E.~G.\ 1987, \apjl, 318, L39 

\bibitem[Falomo et al.(1991)]{falomo91} Falomo, R., Giraud, E., 
Melnick, J., Maraschi, L., Tanzi, E.~G., \& Treves, A.\ 1991, \apjl, 380, 
L67 

\bibitem[Falomo et al.(1993)]{falomo93} Falomo, R., Bersanelli, 
M., Bouchet, P., \& Tanzi, E.~G.\ 1993, \aj, 106, 11 

\bibitem[Falomo, Scarpa, \& Bersanelli(1994)]{falomo94} Falomo, 
R., Scarpa, R., \& Bersanelli, M.\ 1994, \apjs, 93, 125 

\bibitem[Falomo et al.(1995)]{falomo95} Falomo, R., Pesce, 
J.~E., \& Treves, A.\ 1995, \apjl, 438, L9 

\bibitem[Falomo(1996)]{falomo96} Falomo, R.\ 1996, \mnras, 283, 
241 

\bibitem[Falomo \& Kotilainen (1999)]{falomo99} Falomo, R. \& Kotilainen, J.K. 1999, \aap, 352, 85

\bibitem[Falomo et al.(2000)]{falomo00a} Falomo, R., Scarpa, R., 
Treves, A., \& Urry, C.~M.\ 2000, \apj, 542, 731 

\bibitem[Falomo \& Ulrich(2000)]{falomo00} Falomo, R.~\& Ulrich, 
M.-H.\ 2000, \aap, 357, 91

\bibitem[Feigelson et al.(1986)]{feigelson86} Feigelson, E.~D., et 
al.\ 1986, \apj, 302, 337 

\bibitem[Filippenko et al.(1986)]{filippenko86} Filippenko, A.~V., 
Djorgovski, S., Spinrad, H., \& Sargent, W.~L.~W.\ 1986, \aj, 91, 49 

\bibitem[Fosbury \& Disney(1976)]{fosbury76} Fosbury, R.~A.~E., 
\& Disney, M.~J.\ 1976, \apjl, 207, L75 

\bibitem[Gioia et al.(1990)]{gioia90} Gioia, I.~M., Maccacaro, 
T., Schild, R.~E., Wolter, A., Stocke, J.~T., Morris, S.~L., \& Henry, 
J.~P.\ 1990, \apjs, 72, 567 

\bibitem[Giommi et al.(2005)]{SS} Giommi, P., Piranomonte, 
S., Perri, M., \& Padovani, P.\ 2005, \aap, 434, 385 

\bibitem[Giovannini et al.(2004)]{giovannini04} Giovannini, G., 
Falomo, R., Scarpa, R., Treves, A., \& Urry, C.~M.\ 2004, \apj, 613, 747 

\bibitem[Heidt et al.(1999)]{heidt99} Heidt, J., Nilsson, K., 
Sillanp{\" a}{\" a}, A., Takalo, L.~O., \& Pursimo, T.\ 1999, \aap, 341, 
683 

\bibitem[Heidt et al.(2004)]{heidt04} Heidt, J., Tr{\" o}ller, 
M., Nilsson, K., J{\" a}ger, K., Takalo, L., Rekola, R., \& Sillanp{\" 
a}{\" a}, A.\ 2004, \aap, 418, 813 

\bibitem[Hook et al.(2003)]{hook03} Hook, I.~M., Shaver, 
P.~A., Jackson, C.~A., Wall, J.~V., \& Kellermann, K.~I.\ 2003, \aap, 399, 
469

\bibitem[Horan et al.(2004)]{horan04} Horan, D., et al.\ 2004, 
\apj, 603, 51 
 
\bibitem[Horne (1986)]{horne86} Horne, K., \ 1986, \pasp, 98, 609

\bibitem[Kotilainen et al.(1998)]{kotilainen98} Kotilainen, J.~K., 
Falomo, R., \& Scarpa, R.\ 1998, \aap, 336, 479 

\bibitem[Kotilainen \& Falomo(2004)]{kotilainen04} Kotilainen, 
J.~K., \& Falomo, R.\ 2004, \aap, 424, 107 

\bibitem[Kinney et al.(1996)]{kinney96} Kinney, A.~L., Calzetti, 
D., Bohlin, R.~C., McQuade, K., Storchi-Bergmann, T., \& Schmitt, H.~R.\ 
1996, \apj, 467, 38 

\bibitem[Landt et al.(2001)]{landt01} Landt, H., Padovani, P., 
Perlman, E.~S., Giommi, P., Bignall, H., \& Tzioumis, A.\ 2001, \mnras, 
323, 757 

\bibitem[Laurent-Muehleisen et al.(1998)]{rgb} 
Laurent-Muehleisen, S.~A., Kollgaard, R.~I., Ciardullo, R., Feigelson, 
E.~D., Brinkmann, W., \& Siebert, J.\ 1998, \apjs, 118, 127

\bibitem[Londish et al.(2002)]{londish02} Londish, D., et al.\ 
2002, \mnras, 334, 941 

\bibitem[March{\~a} et al.(1996)]{marcha96} March{\~a}, M.~J.~M., 
Browne, I.~W.~A., Impey, C.~D., \& Smith, P.~S.\ 1996, \mnras, 281, 425 

\bibitem[Margon \& Jacoby(1984)]{margon84} Margon, B., \& 
Jacoby, G.~H.\ 1984, \apjl, 286, L31 

\bibitem[Nilsson et al.(2003)]{nilsson03} Nilsson, K., Pursimo, T., Heidt, J.,
  Takalo, L.~O., Sillanp{\"a}{\"a}, A., \& Brinkmann, W.\ 2003 \aap, 400, 95

\bibitem[Oke(1990)]{oke90} Oke, J.~B.\ 1990, \aj, 99, 1621

\bibitem[Owen \& Muffson(1977)]{owen77} Owen, F.~N., \& 
Muffson, S.~L.\ 1977, \aj, 82, 776 

\bibitem[Padovani \& Giommi(1995)]{padovani95} Padovani, P., \& 
Giommi, P.\ 1995a, \mnras, 277, 1477 

\bibitem[Pian et al.(1996)]{pian96} Pian, E., Falomo, R., 
Ghisellini, G., Maraschi, L., Sambruna, R.~M., Scarpa, R., \& Treves, A.\ 
1996, \apj, 459, 169 

\bibitem[Pian et al.(2005)]{pian05} Pian, E., Falomo, R., \& 
Treves, A.\ 2005, \mnras, 361, 919 
 
\bibitem[Penton et al.(2004)]{penton04} Penton, S.~V., Stocke, 
J.~T., \& Shull, J.~M.\ 2004, \apjs, 152, 29  

\bibitem[Perlman et al.(1996)]{perlman96} Perlman, E.~S., et al.\ 
1996, \apjs, 104, 251 

\bibitem[Pesce et al.(1994)]{pesce94} Pesce, J.~E., Falomo, R., 
\& Treves, A.\ 1994, \aj, 107, 494 

\bibitem[Pesce et al.(1995)]{pesce95} Pesce, J.~E., Falomo, R., 
\& Treves, A.\ 1994, \aj, 110, 1554

\bibitem[Pesce et al.(1997)]{pesce97} Pesce, J.~E., et al.\ 
1997, \apj, 486, 770 

\bibitem[Rector \& Stocke(2001)]{rector01} Rector, T.~A., \& 
Stocke, J.~T.\ 2001, \aj, 122, 565 

\bibitem[Richards et al.(2002)]{richards02} Richards, G.~T., et 
al.\ 2002, \aj, 123, 2945 

\bibitem[Sbarufatti et al.(2005a)]{SI} Sbarufatti, B., Treves, A.,
Falomo, R., Heidt, J., Kotilainen, J., Scarpa, R.\ 2005a , \aj, 129, 599
(S05a)

\bibitem[Sbarufatti et al.(2005b)]{SIII} Sbarufatti, B., 
Treves, A., \& Falomo, R.\ 2005b, \apj, 635, 173 (S05b)
 
\bibitem[Sbarufatti et al.(2006)]{SII} Sbarufatti, B., Treves, A.,
Falomo, R., Heidt, J., Kotilainen, J., Scarpa, R.\ 2006 , \aj, in press, 
astro-ph/0601506 (S06)

\bibitem[Scarpa et al.(1995)]{scarpa95} Scarpa, R., Falomo, R., 
\& Pian, E.\ 1995, \aap, 303, 730 

\bibitem[Scarpa et al.(1999)]{scarpa99} Scarpa, R., Urry, C.~M., 
Falomo, R., \& Treves, A.\ 1999, \apj, 526, 643 

\bibitem[Scarpa et al.(2000a)]{scarpa00a} Scarpa, R., Urry, C.~M., 
Falomo, R., Pesce, J.~E., \& Treves, A.\ 2000, \apj, 532, 740 

\bibitem[Schachter et al.(1993)]{schachter93} Schachter, J.~F., et 
al.\ 1993, \apj, 412, 541 

\bibitem[Schlegel, Finkbeiner, \& Davis(1998)]{schlegel98} 
Schlegel, D.~J., Finkbeiner, D.~P., \& Davis, M.\ 1998, \apj, 500, 525 

\bibitem[Sowards-Emmerd et al.(2005)]{sowards05} Sowards-Emmerd, D.,
Romani, R.~W., Michelson, P.~F., Healey, S.~E., Nolan, P.~L. \ 2005, \apj, 626, 95

\bibitem[Stecker et al.(1996)]{stecker96} Stecker, F.~W., de 
Jager, O.~C., \& Salamon, M.~H.\ 1996, \apjl, 473, L75 

\bibitem[Stickel \& K{\"u}hr(1993)]{stickel93} Stickel, M., \& K{\"u}hr, H.\ 
1993, \aaps, 100, 395

\bibitem[Stocke et al.(1985)]{stocke85} Stocke, J.~T., Liebert, 
J., Schmidt, G., Gioia, I.~M., Maccacaro, T., Schild, R.~E., Maccagni, D., 
\& Arp, H.~C.\ 1985, \apj, 298, 619 

\bibitem[Tody(1986)]{tody86} Tody, D.\ 1986, \procspie, 627, 
733 

\bibitem[Tody(1993)]{tody93} Tody, D.\ 1993, ASP Conf.~Ser.~ 
52: Astronomical Data Analysis Software and Systems II, 2, 173 

\bibitem[Urry et al.(2000)]{urry00} Urry, C.~M., Scarpa, R., 
O'Dowd, M., Falomo, R., Pesce, J.~E., \& Treves, A.\ 2000, \apj, 532, 816 

\bibitem[V{\'e}ron-Cetty \& V{\'e}ron(1993)]{veron93} V{\'e}ron-Cetty, 
M.-P., \& V{\'e}ron, P.\ 1993, \aaps, 100, 521 

\bibitem[Villata et al.(2000)]{villata00} Villata, M., Raiteri, 
C.~M., Popescu, M.~D., Sobrito, G., De Francesco, G., Lanteri, L., \& 
Ostorero, L.\ 2000, \aaps, 144, 481 

\bibitem[Weiler \& Johnston(1980)]{weiler80} Weiler, K.~W., \& 
Johnston, K.~J.\ 1980, \mnras, 190, 269 
 
\bibitem[White et al.(2000)]{white00} White, R.~L., et al.\ 
2000, \apjs, 126, 133 

\bibitem[Wills \& Wills(1976)]{wills76} Wills, D., \& Wills, 
B.~J.\ 1976, \apjs, 31, 143 

\bibitem[Wills \& Wills(1979)]{wills79} Wills, B.~J., \& Wills, 
D.\ 1979, \apjs, 41, 689 

\bibitem[Woo et al.(2005)]{woo05} Woo, J.-H., Urry, M.~C., 
van der Marel, R.~P., Lira, P., Maza, J., \ 2005, \apj, 631, 762

\bibitem[Wright et al.(1998)]{wright98} Wright, S.~C., McHardy, 
I.~M., Abraham, R.~G., \& Crawford, C.~S.\ 1998, \mnras, 296, 961 

\bibitem[Wurtz et al.(1996)]{wurtz96} Wurtz, R., Stocke, J.~T., 
\& Yee, H.~K.~C.\ 1996, \apjs, 103, 109 

\end{thebibliography}
\end{document}